\documentclass[manuscript]{acmart} 

\renewcommand\footnotetextcopyrightpermission[1]{} 
\AtBeginDocument{%
  \providecommand\BibTeX{{%
    \normalfont B\kern-0.5em{\scshape i\kern-0.25em b}\kern-0.8em\TeX}}}

\usepackage{multirow}
\usepackage{booktabs}
\usepackage{makecell}
\usepackage{colortbl}
\usepackage{colortbl}
\usepackage{xcolor}
\usepackage{soul}
\usepackage{cleveref}

\newcommand{\PRIMEproject}{PRIME project } 
\newcommand{\PRIMEprojectdot}{PRIME project. } 
\newcommand{\PRIMEprojects}{PRIME project's } 
\newcommand{\PRIME}{PRIME } 
\newcommand{\grant}{URKI’s Engineering and Physical Sciences Research Council (EPSRC) grant numbers EP/W03235X/1, EP/W032333/1, EP/W032341/1, EP/ W032058/1, EP/W032082/1 under the Protecting Minority Ethnic Communities Online (PRIME) project. }

\newcommand{\fourcities}{Glasgow, Bradford, Manchester and Tower Hamlets }

\begin{document}

\title{A Unifying Bias-aware Multidisciplinary Framework for Investigating Socio-Technical Issues}

\author{Sacha Hasan}
\affiliation{%
  \institution{School of Energy, Geoscience, Infrastructure and Society, Heriot-Watt University}
  \city{Edinburgh}
  \country{United Kingdom}
}
\email{sacha.hasan@hw.ac.uk}
\orcid{0000-0003-1703-1017}
\authornote{Both authors contributed equally to this research} 

\author{Mehdi Rizvi}
\affiliation{%
  \institution{School of Mathematical and Computer Sciences, Heriot-Watt University}
  \city{Edinburgh}
  \country{United Kingdom}
}
\email{s.rizvi@hw.ac.uk}
\orcid{0000-0001-8386-5779}
\authornotemark[1]

\author{Yingfang Yuan}
\affiliation{%
  \institution{School of Mathematical and Computer Sciences, Heriot-Watt University}
  \city{Edinburgh}
  \country{United Kingdom}
}
\email{y.yuan@hw.ac.uk}
\orcid{0000-0002-8925-9267}

\author{Kefan Chen}
\affiliation{%
  \institution{School of Mathematical and Computer Sciences, Heriot-Watt University}
  \city{Edinburgh}
  \country{United Kingdom}
}
\email{kc2039@hw.ac.uk}
\orcid{0009-0001-5726-466X}

\author{Lynne Baillie}
\affiliation{%
  \institution{School of Mathematical and Computer Sciences, Heriot-Watt University}
  \city{Edinburgh}
  \country{United Kingdom}
}
\email{L.Baillie@hw.ac.uk}
\orcid{0000-0002-2514-5981}

\author{Wei Pang}
\affiliation{%
  \institution{School of Mathematical and Computer Sciences, Heriot-Watt University}
  \city{Edinburgh}
  \country{United Kingdom}
}
\email{w.pang@hw.ac.uk}
\orcid{0000-0002-1761-6659}

\renewcommand{\shortauthors}{Hasan et al.}

\begin{abstract}

This paper aims to bring together the disciplines of social science (SS) and computer science (CS) in the design and implementation of a novel multidisciplinary framework for systematic, transparent, ethically-informed, and bias-aware investigation of socio-technical issues. For this, various analysis approaches from social science and machine learning (ML) were applied in a structured sequence to arrive at an original methodology of identifying and quantifying objects of inquiry. A core feature of this framework is that it highlights where bias occurs and suggests possible steps to mitigate it. This is to improve the robustness, reliability, and explainability of the framework and its results. Such an approach also ensures that the investigation of socio-technical issues is transparent about its own limitations and potential sources of bias.
To test our framework, we utilised it in the multidisciplinary investigation of the online harms encountered by minoritised ethnic (ME) communities when accessing and using digitalised social housing services in the UK. We draw our findings from 100 interviews with ME individuals in four cities across the UK to understand ME vulnerabilities when accessing and using digitalised social housing services. In our framework, a sub-sample of interviews focusing on ME individuals residing in social housing units were inductively coded. This resulted in the identification of the topics of discrimination, digital poverty, lack of digital literacy, and lack of English proficiency as key vulnerabilities of ME communities. Further ML techniques such as Topic Modelling and Sentiment Analysis were used within our framework where we found that Black African communities are more likely to experience these vulnerabilities in the access, use and outcome of digitalised social housing services.


\end{abstract}


   


\maketitle

\section{Introduction}

Multidisciplinary research approaches are becoming increasingly popular in investigating complex socio-technical questions as they enable “meaningful knowledge co-production through integrative and participatory processes that bring together diverse actors, disciplines, and knowledge bases” ~\cite{thompson2017scientist}. 
Furthermore, a large amount of machine learning (ML) techniques are being applied in various disciplines (e.g., chemistry\cite{artrith2021best}, biology\cite{tarca2007machine}, social science \cite{hindman2015building}), in an interdisciplinary fashion.
The applications of machine learning to social science data can facilitate uncovering novel concepts, measuring the frequency of these concepts, evaluating causal relationships, discovering trends and patterns, and generating models to support decision-making for real-life scenarios.

However, despite the generally positive attitude towards multidisciplinary research, there is a number of challenges which arise when individuals with divergent philosophies, expertise, expectations, and conflicts work in a multidisciplinary setting. 
This can include additional time and effort to support communication around roles and responsibilities; a hazard to conducting research and conceptualising its findings within different disciplinary contexts, and within traditional social and institutional structures~\cite{dehart2017team, thompson2017scientist}. 
Despite this conceptualisation of the approach and the identification of its benefits and challenges, it is important to note that there is no widely agreed definition or framework to shape a multidisciplinary approach to research. 

The speed of producing and storing data has soared to unprecedented rate in social sciences~\cite{grimmer2015we}. 
These datasets are “big” not just in their size and the speed at which they accumulate, but also in their complexity and variety~\cite{hindman2015building}. 
However, and despite the evident benefit, it is important to recognise that “big data” alone is insuﬃcient for solving social problems, mainly due to measurement error, and other sources of bias which often make social scientists sceptical~\cite{grimmer2015we}. 
Thus, to identify a social phenomenon, ML analysis requires thoughtful measurement, careful research design, and the creative deployment of statistical techniques - blending social science with computer science~\cite{grimmer2015we,patty2015analyzing}. 

Having said that and despite this era of data abundance, datasets can come in all sizes; big, small and medium-sized. Furthermore, and this is especially relevant to social sciences, datasets are not always survey based, but more often include a wide range of qualitative data that is recorded in more than one format. This is in addition to the curse of the multidimensional nature of social phenomena often recorded in everyday experiences which often hammered social researchers when attempting to better quantify the object of inquiry and to accurately predict human behaviour and social structures~\cite{hindman2015building}. 
These limitations to the traditional social research approaches to data handling and the shortcomings of classical computational data analysis have invited a shift in how we attempt to understand and quantify social \textit{object of inquiry}. One approach to this is the use of ML to extract meaning, discover new concepts, measure the prevalence of those concepts, assess causal effects, and make predictions from big datasets and - or rather especially relevant to social research – small and medium-sized ones~\cite{grimmer2015we,hindman2015building}. 

Nonetheless, it is important to mention here that the applications of ML in social research are often developed with limited input from experts outside of computational fields. Thus, social scientists are increasingly recognised to have considerable expertise to contribute to the development of ML applications for human-generated data and their analytic practices to arrive at more human-centred ML modules~\cite{chen2018using}. 

This revolution in the practice of social science has already recorded advances in the areas of genomics and medicine~\cite{rahal2022rise}. 
Thus, there is a potential to urge rethinking of other areas of social science research to ``a more sequential, interactive, and ultimately inductive approach to inference''~\cite{grimmer2021machine,hofman2021integrating}. 
The combination of human attention to certain aspects of the object of inquiry and the power of algorithms to automate other aspects of the same object of inquiry would allow a special synergy, capable of overcoming the limitations of each of the disciplines when used alone~\cite{lundberg2022researcher}. This can contribute to the reduction of bias from both sources i.e., human bias and the bias introduced by the algorithms and data. 

In this context, our multidisciplinary framework highlights that
ML techniques are increasingly accessible and capable of yielding new discoveries for social research. They amplify the social researchers' inductive coding, identify patterns, summarise complex data, relax statistical assumptions, and, thus, allow better predictive accuracy~\cite{hindman2015building}. 
More importantly, our framework allows a robust approach to data handling that is more likely to be replicated to fit different areas of investigation compared to the traditional methods often adopted in empirical social research~\cite{hindman2015building}. This is while highlighting various  potential sources of bias often introduced during research. Being aware of this bias is important to ensure informed decisions are made when ML techniques are applied to socio-technical problems. 

\section{Related work}
\subsection{Bias in Collecting Data, During Interaction with Data, and in Data Analysis Methods}\label{sourcesofbias}

Research is prone to various types of bias. Researchers could affirm their biases during the research design, data collection and data analysis processes~\cite{pannucci2010identifying}. In the context of social sciences, bias is defined as the tendency which prevents unprejudiced consideration of the subject of inquiry \cite{dictionarycom2012)}. 
This often occurs when ``systematic error is introduced into sampling or testing by selecting or encouraging one outcome or answer over others''~\cite{MerriamWebster2023}. 
Within this, it is not possible to consider bias as a dichotomous variable – limited to the question of ``is bias present or not?'' Alternatively, researchers should address the ``type'' and the ``degree'' of bias, and how this impacts the research findings and conclusions~\cite{pannucci2010identifying}. 

\begin{figure}[htb]
    \centering
    \includegraphics[width=.7\textwidth]{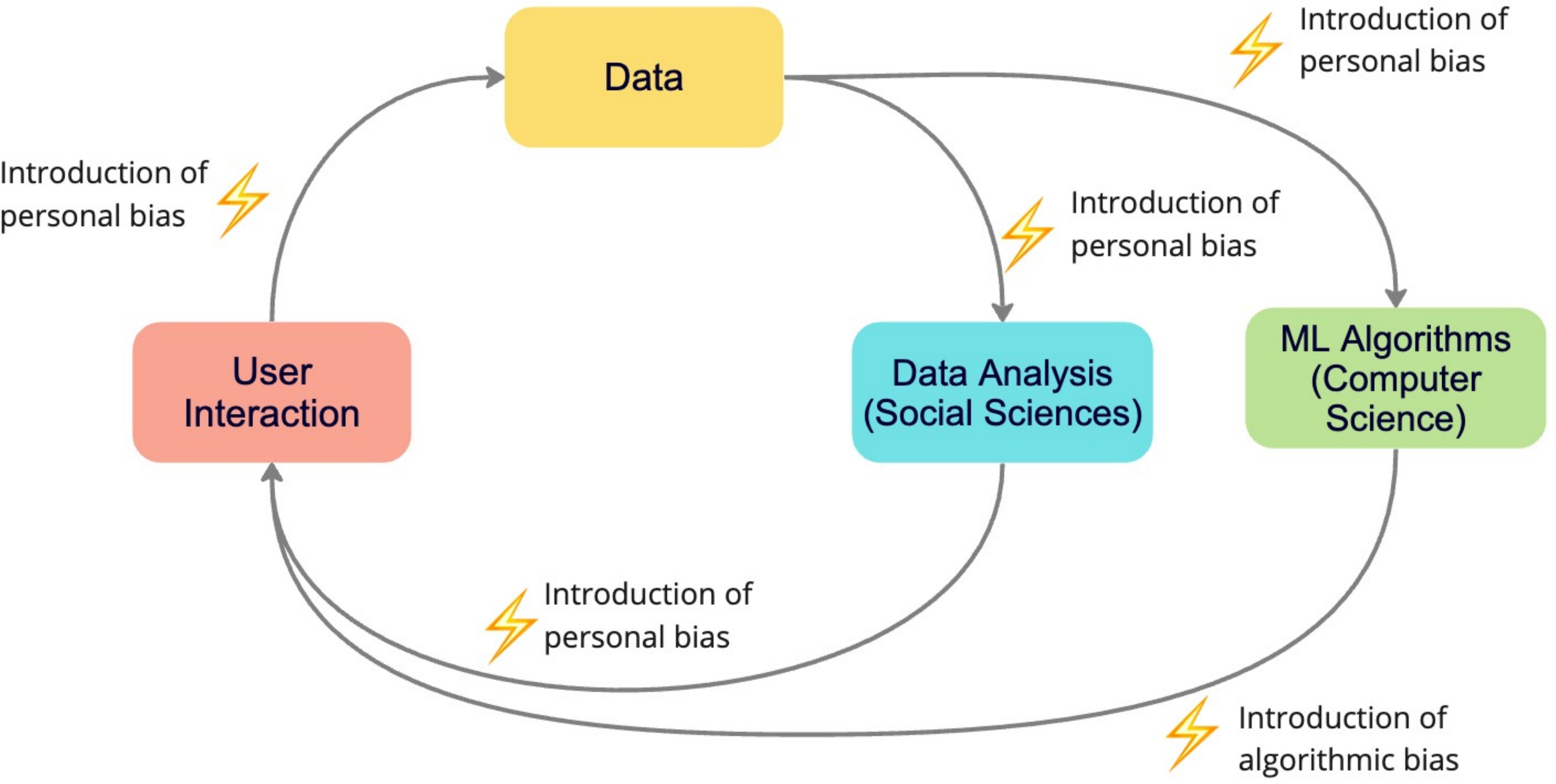}
    \caption{The cyclic nature of data, interaction with data, and, data analysis methods/algorithms feeding into each other and building up on biases}
    \Description{The cyclic nature of data, interaction with data, and, data analysis methods/algorithms feeding into each other and building up on biases}
    \label{fig:cycleofbias}
\end{figure}

Other sources of biases are from the data itself and the algorithms used to collect or process the data. If the algorithms are trained on biased data, the outputs of such algorithms are likely to be inherently biased ~\cite{mehrabi2021survey}. All these sources of bias feed into each other in a cyclic manner in the various stages of research design, data collection, and data analysis. See Fig~\ref{fig:cycleofbias} for an illustration.

It is important to accept that bias is inevitable in research. Therefore, various strategies can be employed to mitigate it. One practice is maintaining reflexivity throughout the interpretation process. This involves researchers reflecting on their own beliefs, values, and experiences, recognising how these may impact and shape the research process and consciously working to minimise their influence~\cite{bergerR2015now}. 
Another practice is to employ data triangulation, where multiple sources and types of data are used to verify the findings and ensure their validity~\cite{denzin2017research}. 

\subsection{Using Machine Learning Techniques for Social or Sociotechnical Research}

\subsubsection{Topic modelling}
Topic modelling is a type of statistical machine learning models that is utilised to examine clusters of topics that co-occur together within a collection of texts. This approach is often employed to comprehend recurring themes and patterns \cite{wesslen2018computer}. Each topic is defined by a set of words. There are many related applications in social science \cite{pandur2020topic, valdez2018topic, baranowski2023social, wesslen2018computer}. We hypothesise that topic modelling could aid in extracting high-level biases or bias-related information from interview transcripts.  

Latent Dirichlet Allocation (LDA) \cite{jelodar2019latent} is a widely used topic modelling technique in natural language processing and machine learning. LDA is a probabilistic generative model that helps identify topics within a collection of documents. \cite{baranowski2023social} used LDA to investigate the diversity and complexity of social welfare issues based on the published papers from 2000 to 2020. The research \cite{baranowski2023social} found that the concepts of welfare can carry varied interpretations for different individuals. Latent Semantic Analysis (LSA)~\cite{landauer1998introduction} employs a Bag-of-words model~\cite{zhang2010understanding}, resulting in generating of a term-document matrix that captures term occurrences within documents. In this matrix, rows represent terms, while columns represent documents. LSA extracts latent topics by decomposing the matrix through Singular Value Decomposition (SVD)~\cite{valdez2018topic}. LSA has been used to analyse the transcripts of the 2016 US presidential debates between Hillary Clinton and Donald Trump~\cite{valdez2018topic}. The study found the topics reflected nuanced differences between the two, and emphasised the importance of topic modelling in social science. Furthermore, Contrastive Explanation (CorEx)~\cite{gallagher2017anchored} is a semi-supervised topic modelling method; it does not assume an underlying generative model such as LDA, and instead it learns maximally informative topics through an information theoretic framework. The approach also allows users to add anchor words to improve topic separability and representation. The research~\cite{luo2023hiking} utilised CorEx to analyse hikers' interests and concerns regarding the experience attributes during different seasons, and the study found that certain attributes exhibit greater sensitivity during specific seasons. Additionally, as the development of large language models, BERTopic~\cite{grootendorst2022bertopic} is an advanced topic modelling technique that leverages pre-trained transformer-based models like BERT (Bidirectional Encoder Representations from Transformers) for more accurate and context-aware topic modelling. Traditional topic modelling methods like LDA are limited in capturing the semantic nuances of text data. BERTopic, on the other hand, applies BERT embedding to represent each document in a corpus and then employs dimensionality reduction and clustering techniques to discover meaningful topics. The research \cite{sharifian2022analysing} applied BERTopic to analyse longitudinal social science questionnaires to uncover semantic changes and evolution of the topics and identified instances of mislabelled question tags.

\subsubsection{Explainable Artificial Intelligence (XAI)}
In ML algorithms, especially deep neural networks, it is difficult for the models to explain how their prediction results are derived, and this is also called black box problems~\cite{adadi2018peeking}.  However, as AI is increasingly used in public scenarios, people's concerns and criticisms about the ethics, safety, transparency and trustworthiness of AI have become more serious~\cite{trocin2021responsible}. XAI aims to address those concerns by making AI systems more understandable to users to increase reliability and trust~\cite{maltbie2021xai}. Currently, XAI has already attracted much attention from different domains. At the same time, many XAI techniques have also been developed and applied in both academic and industrial fields. Based on the different explainable stages, the current XAI techniques can be divided into three distinct categories: data explainability, feature-based techniques, and example-based techniques~\cite{dwivedi2023explainable}. Some useful and common technologies include data visualisation with dimensionality reduction methods, Local Interpretable Model-Agnostic Explanations (LIME)~\cite{ribeiro2016should}, SHapley Additive exPlanations (SHAP)~\cite{lundberg2017unified}, and Kernel SHAP~\cite{lundberg2017unified}. In the study of academic performance \cite{sargsyan2020explainable}, LIME with $k$-means clustering has been applied to identify groups of students with similar academic attainment indicators. Furthermore, both LIME and SHAP were employed to explain the results obtained from random forest and deep learning in the study of credit card fraud \cite{ji2021explainable}. In the study, the explanation results were evaluated through a quantitative survey, indicating that XAI explanations have the potential to marginally enhance users' perceptions of the system's reasoning capabilities. Additionally, SHAP was applied  to investigate the factor of influencing citations based on the CatBoost model, the outcomes demonstrated that the year has a substantial impact on citation but does not influence the priority factor \cite{ha2022explainable}. 

\subsubsection{Latent Class Analysis (LCA)}
Latent Class Analysis (LCA) is a statistical method based on mixture model which is often used to detect potential or unobserved heterogeneity in samples~\cite{hagenaars2002applied}. According to the response patterns of observed variables, LCA could identify potential subgroups in a sample ~\cite{muthen2000integrating}. Because LCA is a person-centred mixture model, it is widely used in sociology and statistics to interpret and identify different subgroups in a population that often share certain external characteristics from data~\cite{weller2020latent}. Currently, LCA's application areas span health~\cite{grant2020use, mori2020using}, energy~\cite{bardazzi2023energy}, and social fields~\cite{martinez2020cyberbullying}.

In \cite{grant2020use}, 
researchers used clinical judgement and analytical methods to identify the most informative variables and grouped patients using latent class analysis and k-means clustering. Results of the analysis identified seven distinct patient characteristics with significantly different 1-year mortality rates and suggested different care strategies for resource allocation and coordination of care interventions. 
The study \cite{mori2020using} highlights the applications of LCA in a variety of healthcare settings, such as identifying patient subgroups with differential treatment responses and capturing participant preferences, classifying participants into different Urinary Albumin to Creatinine Ratio (UACR) trajectories, resulting in finding associations with adverse cardiac function.
%
Researchers in \cite{bardazzi2023energy} developed a tool using LCA, to characterise energy poverty without the need to arbitrarily define binary cutoffs. The authors highlight the need for a multidimensional approach to measuring energy poverty and discuss the challenges of identifying vulnerable consumers. The paper \cite{martinez2020cyberbullying} discussed the relationship between cyberbullying and social anxiety among Hispanic adolescents. 
There were significant differences in cyberbullying patterns across all social anxiety subscales. Compared with other profiles, students with higher cyberbullying traits scored higher on social avoidance and distress in social situations, as well as lower levels of fear of negative evaluation and distress in new situations.

\subsection{Inductive coding (IC)}
Inductive coding (IC) is a interpretivist constructivist data-driven approach where ``the research begins with an area of study and allows the theory to emerge from the data''~\cite{corbin1990grounded, gioia2013seeking, StraussCorbin1998,chandra2019qualitative}. 
Here, data is in the form of textual units (words, sentences and paragraphs) and a researcher reads, interprets and labels them in order to identify emerging concepts and themes that can be positioned back in literature or used for further analysis. Inductive reasoning is the key strategy that relies on examination, repetition, and comparison to identify patterns (actors, occurrences, relationships) within the raw data by grouping codes of similar components. 
These patterns are what is known as emerging concepts which can then be further categorised into established themes that are often recognised in theory. Further, these describe and organise the possible observations at the minimum, and interpret aspects of the social phenomenon at the maximum~\cite{fereday2006demonstrating}.
IC is often used when investigating areas of limited knowledge as it does not require predetermined concepts or codes from theory~\cite{chandra2019qualitative}. 
This is especially useful when the research attempts to create ideas or theories that are unprecedented.

\section{A Bias-aware Multidisciplinary Framework}

\begin{figure}[htb]
    \centering
    \includegraphics[width=.99\textwidth]{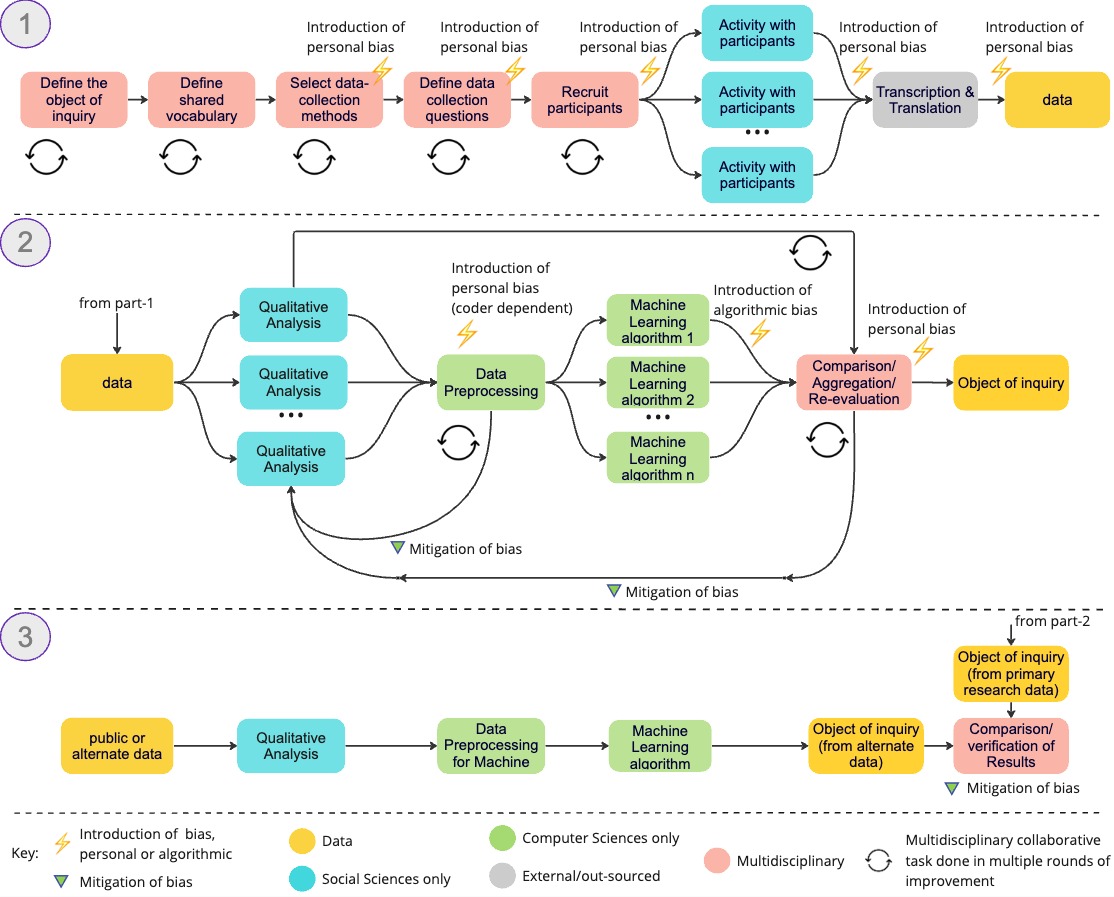}
    \caption{The proposed bias-aware multidisciplinary framework for systematic investigation of socio-technical issues: Part1 (top) refers to the data collection stage of the framework whereas Part2 (bottom) refers to the data analysis stage of the framework.}
    \Description{The proposed bias-aware multidisciplinary framework for systematic investigation of socio-technical issues: Part1 (top) refers to the data collection stage of the framework whereas Part2 (bottom) refers to the data analysis stage of the framework.}
    \label{fig:framework}
\end{figure}

To arrive at a common language to communicate the needs of our research, we invested in critical thinking to develop a number of possibilities to sequence social analysis methods and ML models to allow a meaningful flow of analysis between the different disciplines. We experimented a number of possibilities to test issues of  robustness, reliability, transparency, and explainability of our findings. In addition, we worked to address bias occurrences and how these could be mitigated to inform the sequence of the framework stages and their components. We finally arrived at the proposed bias-aware multidisciplinary framework, shown in Fig~\ref{fig:framework}.  Top part of the figure, labelled as Part-1, refers to the data collection stage of the framework. Part-2 refers to the data analysis stage, whereas Part-3 refers to the verification stage of the framework. Distinct steps of the framework are discussed as follows. 

\subsubsection*{\textbf{Defining the object of inquiry}}
Definition of object of inquiry is to be done collaboratively and repetitively by multidisciplinary researchers. As the object of inquiry is of a socio-technical nature, the social sciences researchers may be more suitable to take a leading role in defining it. 

\subsubsection*{\textbf{Defining a shared vocabulary}}
As the researchers are from very different fields of study, defining a shared vocabulary is a vital step in understanding each other's work better. The same terminology may have different meanings or even have conflicting definitions in different fields. Therefore, this step needs to be done collaboratively and  iteratively 
by multidisciplinary researchers until a consensus is reached. Once a shared vocabulary has been agreed upon, it needs to be used consistently throughout the subsequent stages of the research process. 

\subsubsection*{\textbf{Selecting data collection methods}}
Selection of data collection methods is to be decided collaboratively as well. This could be any method such as interviews, surveys, etc, which can be later translated to quantitative data. The ML researchers can advise the social scientists in making sure that the format of data collected from each chosen method is usable in the ML techniques to be applied at later stages of the research process.

\subsubsection*{\textbf{Defining data collection questions}}
Definition of data collection questions is also to be decided collaboratively, and depending on the data collection methods selected.

\subsubsection*{\textbf{Recruiting participants}}
Once the object of inquiry and the data collection methods and tools have been agreed upon, and all necessary ethics approvals have been gained, the next step is the recruitment of participants. Depending on the object of enquiry, this may need to be done with external partner organisations such as community partners. 

\subsubsection*{\textbf{Data collection}}
This step generates the person-centric qualitative data. This can be done by the researchers individually or in groups, in a linear or parallel fashion. Personal bias would potentially be introduced and the quality of data might not be consistent across researchers.

\subsubsection*{\textbf{Transcription and translation }}
Depending on the need, an additional step of transcription and translation of the collected data may be needed. For example, survey questions or interview questions for different participants could be in different languages, and may need to be translated to one language which is most convenient to the researchers to work with. Care needs to be taken as this step can introduce personal bias during translation. An example of this personal bias is incomplete or low-quality of translation due to the translator not being able to understand a local dialect or they might be biased against speakers of that dialect. 
Transcription may also be needed in cases such as audio data was recorded and textual data is needed to be fed into machine learning algorithms in later stages. 
 
\subsubsection*{\textbf{Qualitative analysis of data}}
    The collected data then goes through a qualitative analysis techniques commonly employed in Social Sciences, for example deductive coding or inductive coding. This is done manually by social scientists individually. Similar to the data collection step, this may also introduce personal bias as the tagging of data may vary among social scientists.

\subsubsection*{\textbf{Data cleaning and pre-processing}}
The collected and tagged data needs to be then cleaned and pre-processed in order to be usable for machine-learning techniques. The inductive tagging done in the previous step may have been done using a software or format which might be difficult to use for the machine-learning algorithm. To avoid additional and manual data cleaning, and pre-processing, computer science or ML researchers can advise social scientist on what kind of data format is best suitable for their needs.

\subsubsection*{\textbf{Machine Learning algorithms}}
Variations of machine learning algorithms could be applied to the pre-processed data, depending on the different data types and tasks. This part of the framework could be changed due to better match the different situations and purposes, for example, using parallel structures or linear structures - more details will be presented in Section 5.2. Due to the biases in the algorithm itself and inherent bias in the datasets used for the training model, machine learning algorithms may introduce biases into their predictions.

\subsubsection*{\textbf{Comparison, aggregation, or re-evaluation}}
Results of different types of ML algorithms or variations of the same types of a ML algorithm can give different results from the same data set. Changing various hyperparameters of the algorithm can yield different results too. Social scientists can work closely with the Computer Science/ML researchers to compare or aggregate different results to decide which of these are more relevant to the object of inquiry and, thus, are more beneficial in solving the socio-technical problem at hand. At times, based on the results of the one ML technique, and views and observations of social scientists during data collection and inductive coding, the ML techniques may be re-evaluated and parameters may be fine tuned. This process of comparison, aggregation, or re-evaluation may go on in an iterative manner until a consensus is reached by the multidisciplinary research team. 

\subsubsection*{\textbf{Verification}}
The final part of the framework is the verification stage. See part-3 of Fig~\ref{fig:framework}. Here the steps can be similar to the part-2, utilising either social sceinces analysis techniques, ML algorithms or both. However, the data being analysed here needs to be either a  publicly available dataset, or an alternate internal data or both. An alternate internal data is simply a different source of data instead of the main data source being analysed in the framework. Comparing the analysis of data and the object of inquiry from the main source with other public or alternate sources of data helps in triangulating the findings, and the verification of results from other sources. 

\subsection{Being bias-aware: Understanding the sources of personal and machine bias}
As highlighted in Section~\ref{sourcesofbias}, people, data and data analysis tools, can all introduce different types of bias. The framework highlights all potential sources of these as well as instances where they can be mitigated.
In the framework shown in Fig~\ref{fig:framework}, the potential sources of bias are highlighted with a lightening bolt icon (\includegraphics[scale=0.25]{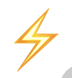}). The opportunities to mitigate bias are highlighted with an inverted triangle icon (\includegraphics[scale=0.3]{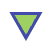}).

Social researchers' knowledge can be utilised to select appropriate and better data sources, machine learning algorithms, and results. The iterative nature of the various steps of the framework provides opportunities to utilise the domain knowledge as well as in triangulating different data sources and results. These opportunities are highlighted with a \includegraphics[scale=0.175]{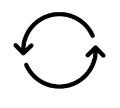} icon in Fig~\ref{fig:framework}.

\subsection{Possible Variants of the Framework}
\begin{figure}[htb]
    \centering
    \includegraphics[width=.9\textwidth]{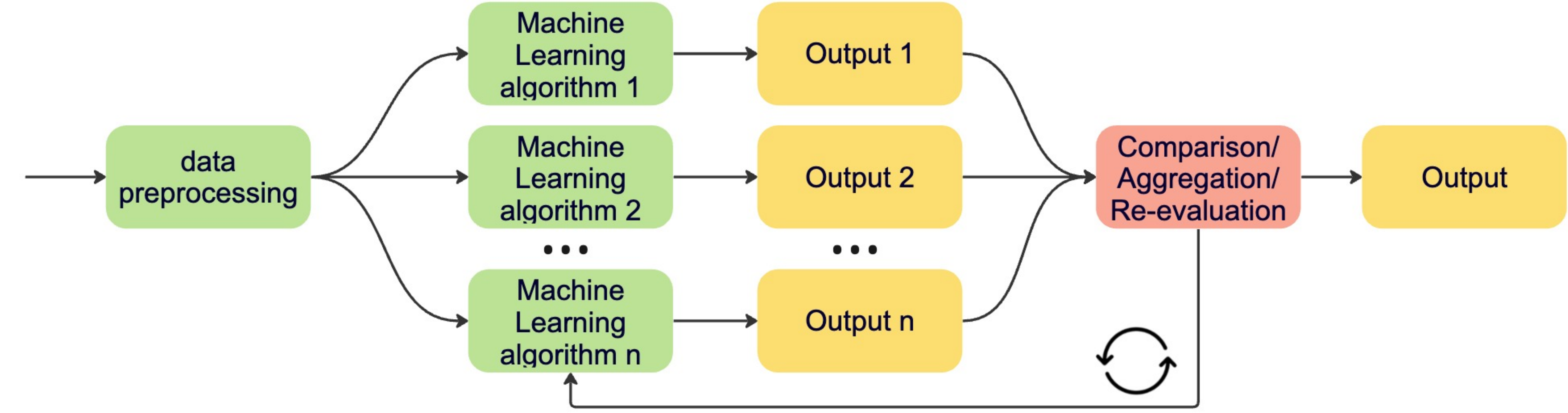}
    \caption{Variant 1: Using different machine learning algorithms in parallel, with multiple possible review cycles. Results of parallel algorithms can be compared, aggregated, selected, and used for fine-tuning before a desired result is achieved. This review cycle need to done in close collaboration with all multidisciplinary research stakeholders.}
     \Description{Parallel variant of framework}
    \label{fig:parallel}
\end{figure}

\begin{figure}[htb]
    \centering
    \includegraphics[width=.9\textwidth]{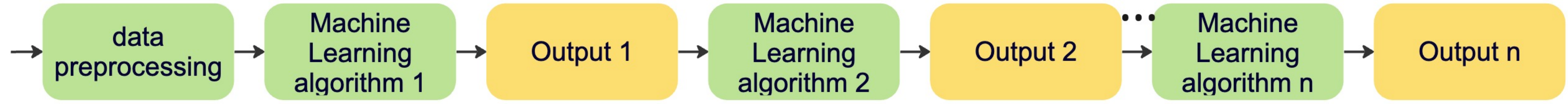}
    \caption{Variant 2: Using different machine learning algorithms in a linear fashion}
    
     \Description{linear variant of framework}
    \label{fig:linear}
\end{figure}

\begin{figure}[htb]
    \centering
    \includegraphics[width=.9\textwidth]{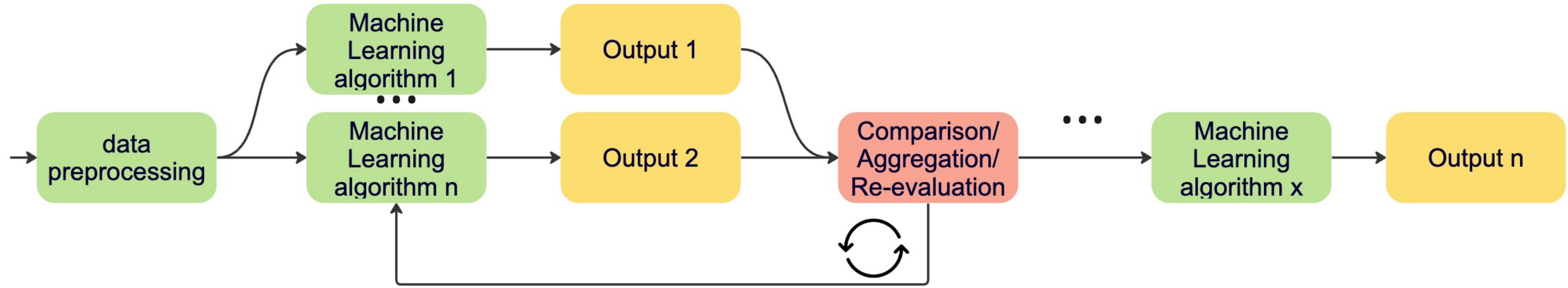}
    \caption{Variant 3: Using different machine learning algorithms in a hybrid modality}
    \label{fig:hybrid}
    
     \Description{Hybrid variant of framework}
\end{figure}

Different ML algorithms or variations of the same ML algorithm could be utilised depending on the data types, tasks and needs. Thus, our framework identifies the use of multiple ML algorithms in a variety of ways: in parallel, one after the other, or in a hybrid fashion. See Fig~\ref{fig:parallel},~\ref{fig:linear} and~\ref{fig:hybrid}, respectively.

Results of the various ML algorithms and/or their variations can be then compared,  aggregated, or used for re-evaluation. This may include fine-tuning the parameters or even discarding the results of an ML algorithm, in favour of another. This can be done in an iterative manner until desired results are achieved. Employing the domain knowledge of social researchers along with triangulation of data and analysis from different sources can help reduce bias. 

\section{Case study: the \PRIMEproject}

\subsection{The context}
The~\PRIMEproject 
brings together interdisciplinary academics from the fields of social science, applied linguistics, computer science and data science. 
\PRIMEproject is driven by the concern that the rapid digitalisation of the key and interconnected services of social housing, health and energy in the UK, accelerated by the pandemic, may exacerbate or replicate inequalities, 
leading to new forms of discrimination against minoritised ethnic (ME) communities. 

\subsection{Applying the bias-aware multidisciplinary framework}

\begin{figure}[htb]
    \centering
    \includegraphics[width=.99\textwidth]{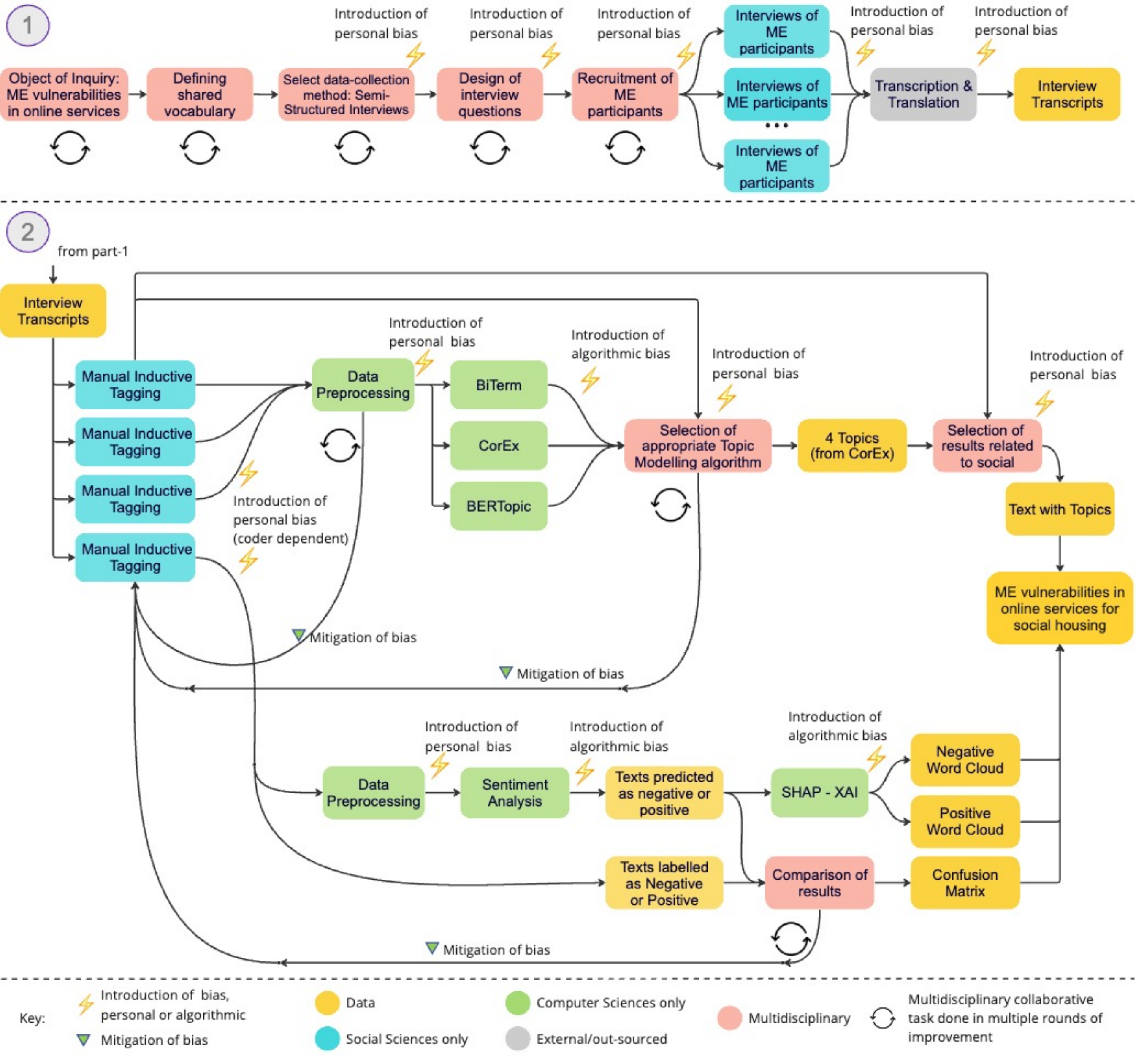}
    \caption{The proposed bias-aware multidisciplinary framework as used with the \PRIME project. Part1 (top) refers to the data collection stage of the \PRIMEproject whereas Part2 (bottom) refers to the data analysis stage of the \PRIME project.}
    \label{fig:primeframework}
\end{figure}

Our proposed framework was utilised in the context of the multidisciplinary investigation of online harms faced by ME groups in the UK in the \PRIME project. See Fig~\ref{fig:primeframework}. The top part of the figure refers to the data collection stage, whereas the bottom part refers to the data analysis stage. 

The object of inquiry, defined through consensus between social scientists and computer science researchers, was "ME vulnerabilities in the access, use, and outcome of digitalised social housing services." Data collection involved semi-structured interviews with 100 ME individuals from \fourcities recruited with the help of ME community organisations. 
Interviews were conducted by four social scientists, resulting in 79 transcripts for analysis to identify ME key vulnerabilities - some participants did not consent to recording and hence didn't have a transcript. As depicted in Fig~\ref{fig:primeframework}, 
a sub-sample of 21 interviews focusing on ME individuals residing in social housing units were inductively coded and then analysed using  Topic Modelling techniques namely BiTerm, CorEx and BERTopic. 

Further ML techniques such as Sentiment Analysis with Explainable AI (XAI) were used on a larger sample of 47 transcripts. These transcripts were selected on the condition that the interviewees used at least one digital service by themselves. Additional data from an internal survey as well as a public survey dataset were analysed using Latent Class Analysis with Exploratory Data Analysis, for the verification of results. These are discussed in detail in Section~\ref{exp_and_results}. 

\subsection{Being Bias-Aware}
In our case study, bias was present in the early stages of research design to investigate a particular group – ME communities, with focus on six specific ethnicities being Chinese, Bangladeshi, Black African, Black Caribbean, Indian, and Pakistani. This could raise the argument of whether the research design was flawed, and whether selection bias was involved. Rather than denying this, the researchers employed the reflexive practice and the triangulation of data by consulting with a wider cohort of the interdisciplinary research team and conducting a systematic literature review to allow a more rigorous process of selection of the research object of inquiry.

Having said that, selection bias was still present in choosing the city case studies, as well as the ME community organisations partners who facilitated the recruitment of interview participants which was done via convenient sampling. Furthermore, the participants' recruitment criteria were flexible allowing different proportions of ages, genders and ethnicities without predetermined balance. Again, this was due to convenience of research resources and time. Furthermore, the design of the data collection tools (the semi-structured interviews for the purpose of this paper) was also subject to researchers’ bias influenced by previous experiences, individual interests and own demographic agendas. To mitigate this, these tools were consulted with the wider interdisciplinary research team in order to reduce prejudices driven by certain individuals or disciplines.

During the data collection stage, interviewers' own bias could have shaped their interactions and responses to participants, which could have impacted the data collected. Attempts to minimise this were used by standardising the researchers’ interaction and consciously maintaining a neutral position from participants’ answers. Researcher bias was also present during the inductive coding stage as their values and experiences may impact how they interpret, label and group text segments of the interviews transcripts under certain themes. To minimise this, Berger’s 
reflexive approach was used to maintain a bias conscious approach during the data analysis process~\cite{bergerR2015now}.

In addition to the bias mitigation practices of reflexivity and triangulation of data, a rigorous approach of interdisciplinary interpretation of data was employed to arrive at a more systematic and less prejudiced findings that are further triangulated with public data sets. 
The cyclic nature of the various steps of the framework helped in utilising the domain knowledge of the social scientists as well as in triangulating different data sources and results. These are highlighted 
in Fig~\ref{fig:primeframework} 
along with various steps which may introduce personal bias or machine learning bias. 

\section{Experiments \& Results}\label{exp_and_results}

This section reports how various ML techniques were used in our case study. The ML techniques used were Topic Modelling techniques (namely BiTerm , BERTopic and CorEx), Sentiment Analysis with XAI, and Latent Class Analysis with Exploratory Data Analysis (for verification of results). These are discussed as follows.

\subsection{Topic modelling}
\label{sec: topicmodelling}
\begin{figure}[hb]
    \centering
    \includegraphics[width=0.95\linewidth]{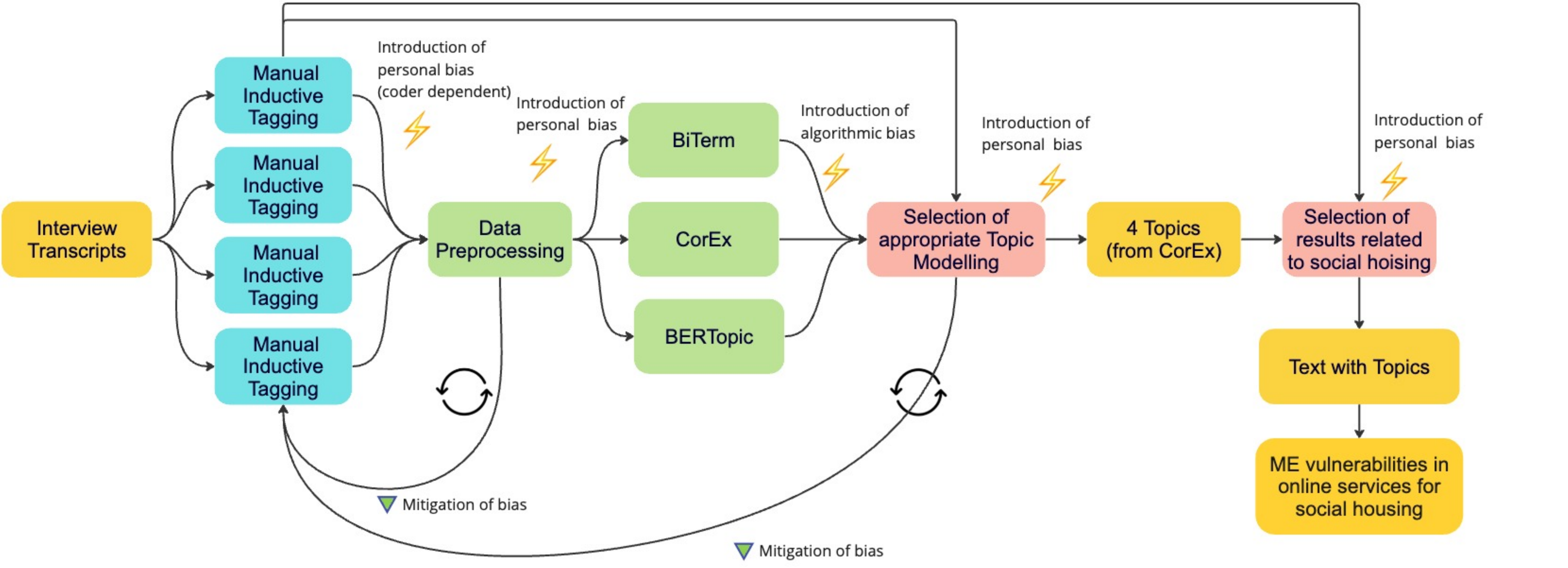}
    \caption{Topic Modelling (including BiTerm , BERTopic , CorEx)}
    \label{fig:ML_TM}
\end{figure}

We suggested to automatically organise interview transcripts into different groups. This was based on the consideration that participants of different ethnicities may have varying positions from and/or experiences with digitalised services. This helped  to identify positive and negative factors and to locate biases. Meanwhile, we worked to classify the text data according to content to systematically investigate different types of answers. For example, text related to different services could be categorised to conduct deeper and fine-grained downstream analysis.
We selected BiTerm ~\cite{yan2013biterm}, CorEx~\cite{gallagher2017anchored}, and BERTopic ~\cite{grootendorst2022bertopic} as the three topic models for our experiments. These three models have their own pros and cons. See Fig.~\ref{fig:ML_TM} for how these were used together in the framework. 

In our comparative experiments, we selected a set of hyperparameters for each topic model. However, the specific hyperparameters to be optimised were chosen based on the designs of the individual models. In other words, different models have different sets of hyperparameters, so we cannot guarantee that the hyperparameter optimisation processes are exactly the same. Nevertheless, we made every effort to design and ensure that they were treated equally. 


For each model, we assigned 100 trials for TPE (Tree-structured Parzen Estimator)~\cite{bergstra2011algorithms} to optimise hyperparameters. The optimised hyperparameter values are shown in Table~\ref{tab:hpsvalues} in \Cref{appendix:topicmodelling}. TPE~\cite{yuan2021systematic} is a approach based on Bayesian optimisation, primarily aimed at finding the optimal hyperparameter configuration in a large search space. By modelling the search space and employing probability density estimation, the TPE algorithm selectively samples within the parameter space, enabling it to identify promising hyperparameter configurations in a relatively short amount of time. BiTerm, CorEx and BERTopic  with the best hyperparameters were respectively optimised with coherence values $-9.5508$, $-2.0025$, $-1.8126$. Coherence UMass \cite{stevens2012exploring} is a metric used to evaluate the quality of topics generated by topic modelling algorithms. It measures the degree of semantic similarity or coherence between words within the same topic. A higher coherence score indicates that the words within a topic are more semantically related, implying that the topic is more coherent and interpretable.

Table \ref{tab:topicwords} shows all results obtained from three optimised topic models. For CorEx, it is noted that the words ``housing'', ``energy'' and ``health'' were used as seed words, taking into account that our interview transcripts and project background mainly focused on these three sectors. Different topic models obtained different numbers of topics. CorEx clearly found four topics respectively related to housing, energy, health, and negative online experiences. In contrast, BiTerm and BERTopic  only found two topics each. See Table~\ref{tab:topicwords} for a list of topics for each topic model, each column represents a topic, and each topic consists of 20 words.

Looking at the results in Table \ref{tab:topicwords}, and in discussion and agreement with the social scientists, the CorEx model was found to be a better fit for the purpose of our case study. It identified a larger number of topics with higher levels of coherence with the 3 digitalised services of ‘housing’, ‘health’ and ‘energy’ while the fourth topic being ‘online’ as investigated in the \PRIMEprojectdot 

\begin{figure}[bt]
    \centering
    \includegraphics[width=0.95\linewidth]{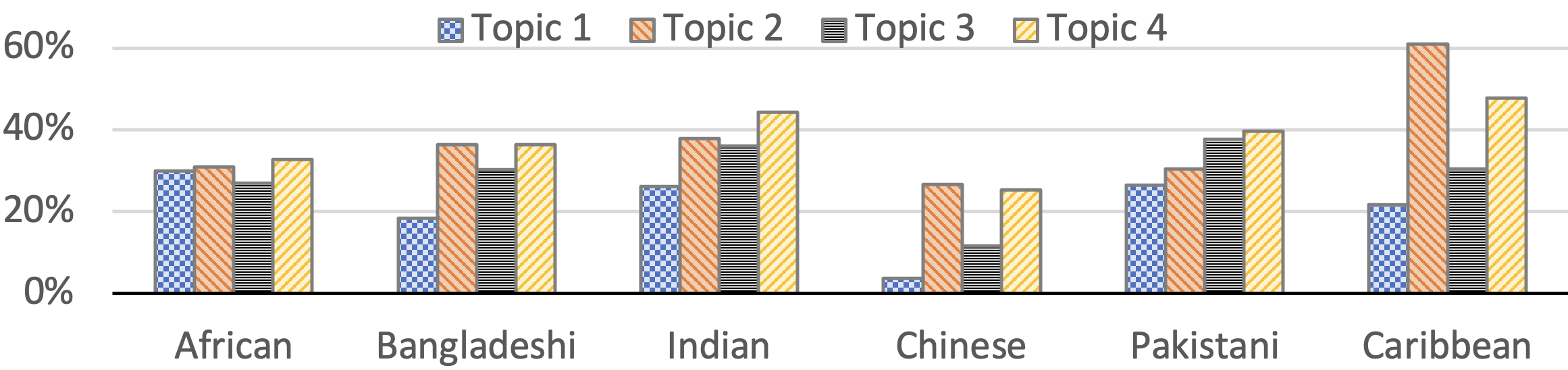}
    \caption{The distributions of topics over different ethnic groups. Topic 1 and Topic 4 relate to Social Housing and online respectively.}
    \label{fig:topicmodelling_percentages}
\end{figure}

For the case study we are examining in this paper, we focused on the topics of social housing (column 1) and online (column 4) to arrive at high-level findings that establish the context of the experiences of the research participants in coherence with the object of inquiry. As topic modelling assigns each transcript to one topic only, we were able to compare which transcripts showed higher relevance to the topics of ‘housing’ and ‘online’. While each transcript relates to one ethnicity, we were able to understand the relevance of ‘housing’ and ‘online’ – being comparatively high or low - to each of the 6 ME groups we investigated in the \PRIMEprojectdot See Fig~\ref{fig:topicmodelling_percentages} for distributions of topics over different ethnic groups.

From this, we can see that the African group has a higher level of relevance to the social housing topic compared to Bangladeshi, Indian, Chinese, Pakistani and Caribbean ethnic groups. This can indicate that this ME group has a greater need, attempt to access, and/or experience of using social housing as an online service. Meanwhile, the analysis has identified that Caribbean and India participants have higher levels of access and use of digitalised services in general without specific indication of which type of service. It is also found that the Chinese group stands out compared to others in housing and energy sectors. These findings help to build awareness around which ME groups have more or less access to digitalised services in comparison to others. This, in turn, supports our understanding of ME vulnerabilities in the access and use of these services including social housing.

\begin{table}[hb]
\centering
\resizebox{\textwidth}{!}{
\begin{tabular}{llll|cl|ll} 
\toprule
\multicolumn{4}{c|}{CorEx}                                               & \multicolumn{2}{c|}{BiTerm } & \multicolumn{2}{c}{BERTopic }  \\
\textcolor{red}{housing}       & energy      & health      & \textcolor{blue}{online}      & meter   & harm              & service     & appointment     \\
\textcolor{red}{social}        & meter       & going       & \textcolor{blue}{experienced} & month   & religious         & internet    & appointments    \\
\textcolor{red}{association}   & supplier    & things      & \textcolor{blue}{negative}    & ring    & afraid            & services    & nhs             \\
\textcolor{red}{council}       & gas         & people      & \textcolor{blue}{experiences} & bill    & ethnicity         & electricity & phone           \\
\textcolor{red}{bidding}       & smart       & lot         & \textcolor{blue}{speak}       & british & cultural          & smartphone  & doctor          \\
\textcolor{red}{accommodation} & electricity & like        & \textcolor{blue}{services}    & flat    & sensitive         & provider    & service         \\
\textcolor{red}{living}        & provider    & really      & \textcolor{blue}{experience}  & sent    & saves             & energy      & services        \\
\textcolor{red}{bid}           & supply      & say         & \textcolor{blue}{saying}      & took    & lack              & getting     & gps             \\
\textcolor{red}{landlord}      & bills       & think       & \textcolor{blue}{harm}        & heating & unfairly          & phone       & doctors         \\
\textcolor{red}{rent}          & british     & feel        & \textcolor{blue}{safe}        & blood   & unsafe            & pay         & app             \\
\textcolor{red}{private}       & tariffs     & way         & \textcolor{blue}{understand}  & meters  & desktop           & online      & medical         \\
\textcolor{red}{house}         & interested  & know        & \textcolor{blue}{somebody}    & paid    & responsive        & account     & prescriptions   \\
\textcolor{red}{apply}         & monitor     & time        & \textcolor{blue}{come}        & six     & navigate          & facebook    & prescription    \\
\textcolor{red}{applying}      & meters      & getting     & \textcolor{blue}{try}         & clock   & religion          & access      & internet        \\
\textcolor{red}{application}   & readings    & look        & \textcolor{blue}{example}     & eight   & relation          & use         & gp              \\
\textcolor{red}{live}          & suppliers   & actually    & \textcolor{blue}{right}       & hour    & altered           & app         & health          \\
\textcolor{red}{flat}          & cut         & good        & \textcolor{blue}{details}     & washing & fear              & email       & email           \\
\textcolor{red}{question}      & electric    & appointment & \textcolor{blue}{mum}         & phoned  & dropin            & meter       & use             \\
\textcolor{red}{contacting}    & change      & bit         & \textcolor{blue}{bank}        & blah    & bullying          & doing       & online          \\
\textcolor{red}{terms}         & compare     & thing       & \textcolor{blue}{using}       & payment & bias              & smart       & okay            \\
\bottomrule
\end{tabular}}
\caption{The results of the three topic models: CorEx, BiTerm , BERTopic . Column 1 and 4 of CorEx were specifically used for our object of inquiry related to social housing.}
\label{tab:topicwords}
\end{table}

Based on these findings, we are able to argue that topic modelling 
helped to establish the understanding of the relevance of each ethnicity to social housing as a service and to access and use of online as a platform. Nonetheless, it is important to note that this stage did not indicate the sentiment of these experiences nor the vulnerabilities of these groups when accessing and using digitalised social housing services. Therefore, we followed the topic modelling with a sentiment analysis. 

\subsection{Sentiment Analysis with Explainable AI}
\label{sec: sentimentanalysis}
An approach to sentiment analysis is to use sentiment-labelled data to fine-tune a pretrained large language model (LLM) such as GPT \cite{brown2020language} or Vicuna \cite{vicuna2023}. 
 Explainable AI (XAI) techniques can then be applied to enhance sentiment analysis 
 by making the sentiment analysis process transparent and understandable. Thus, XAI can address potential issues, improve model performance, and build trust in the generated sentiment predictions.

\begin{figure}[h]
    \centering
    \includegraphics[width=0.95\linewidth]{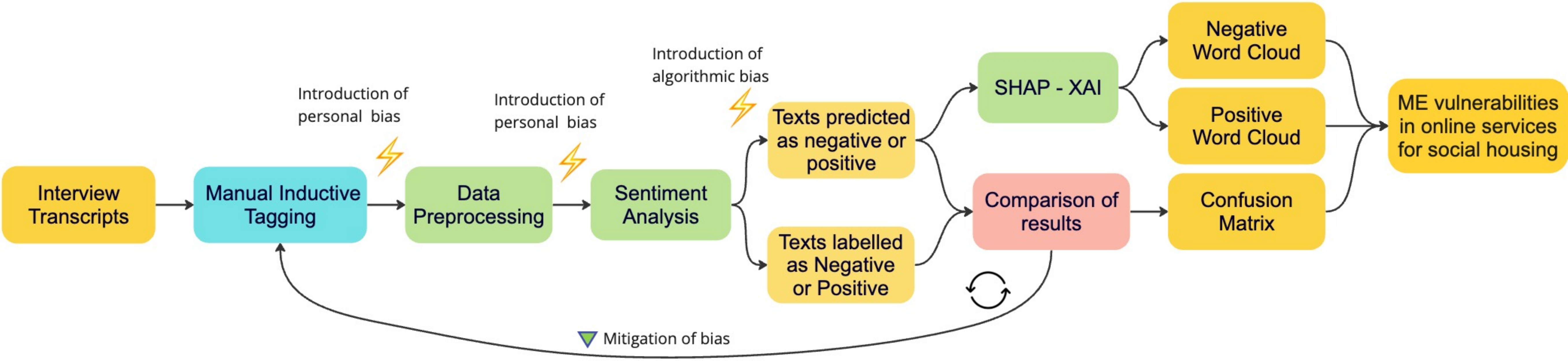}
    \caption{Sentiment Analysis with XAI}
    \label{fig:ML_SAwithXAI}
\end{figure}

The first step was a manual inductive coding (IC) of ME vulnerabilities as described before, and the second was a ML sentiment analysis of the same vulnerabilities. These were done on 21 transcripts. The shortlisting criteria was to select people residing in a social housing unit. The findings obtained from IC informed the subsequent ML analysis. See Fig.~\ref{fig:ML_SAwithXAI} depicting how Sentiment Analysis with XAI was used.

The differences between model predictions and human classifications of ME vulnerabilities are summarised in Table 
 \ref{table: sentiment_analysis}. Here actual values are the numbers of texts that we labelled as positive or negative, whereas predicted values are the numbers of positive/negative predictions made by GPT2-large after it was fine-tuned.  The overall accuracy is 72.881\%, which indicates that approximately 73\% of results from model predictions and human classifications are identical, indicating that the model has a relatively high degree of congruence with human cognition. The negative predictive value is 64\%, which is less than the precision of 79.4\%, indicating that the model's negative classification has room for refinement. However, we believe this was caused by the size of the negative dataset being smaller than the positive dataset. Moreover, in the language domain, we discovered that the accuracy of predictions is 85.7\%. This may indicate that there is a positive relationship between people's ability to convey their positive/negative experience and their language proficiency.

\begin{table}[h]
\centering
\arrayrulecolor[rgb]{0.498,0.498,0.498}
\begin{tabular}{l|lll}
\multicolumn{1}{l}{\textit{~}}             & \textit{~}                                       & \multicolumn{2}{l}{\textit{Actual Values}}                                                           \\ 
\hline
\textit{~}                                 & {\cellcolor[rgb]{0.949,0.949,0.949}}~            & {\cellcolor[rgb]{0.949,0.949,0.949}}Positive (1) & {\cellcolor[rgb]{0.949,0.949,0.949}}Negative (0)  \\
\multirow{2}{*}{\textit{Predicted Values}} & Positive (1)                                     & 27                                               & 9                                                 \\
                                           & {\cellcolor[rgb]{0.949,0.949,0.949}}Negative (0) & {\cellcolor[rgb]{0.949,0.949,0.949}}7            & {\cellcolor[rgb]{0.949,0.949,0.949}}16            \\
\textit{~}                                 & ~                                                & ~                                                & ~                                                
\end{tabular}
\arrayrulecolor{black}
\caption{The confusion matrix summarises the predictions made on the validation data set for our sentiment analysis task over the dimensions of language proficiency and digital literacy}
\label{table: sentiment_analysis}
\end{table}

\begin{figure}[h]
    \centering
    \includegraphics[scale=0.9]{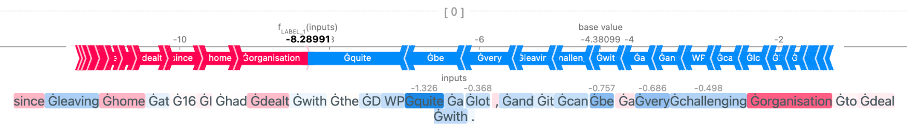}
    \caption{The example of applying SHAP to explain the result of sentiment analysis.}
    \label{fig: shap}
\end{figure}

Fig. \ref{fig: shap} shows an example of SHAP results. It is evident that the words "quiet," "lot," and "challenging," highlighted in blue, reflect a negative sentiment; they convey a strong complaint and may represent vulnerabilities. Subsequently, we iteratively ran the SHAP model (XAI) to extract words from each negative text, and all the collected words were aggregated to generate a word cloud that visually represents the negative terms in an intuitive manner. Based on this, it is possible to consider that negative sentiment analysis can help to investigate the vulnerabilities examined in this experiment. 

\begin{figure}
\centering
\parbox{6cm}{
\includegraphics[width=6cm]{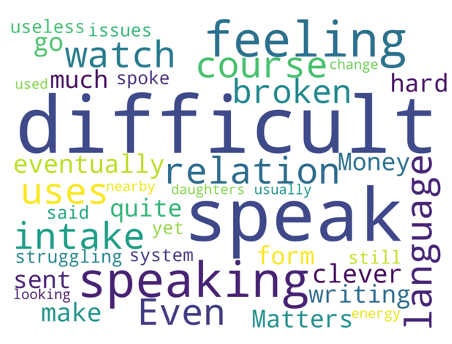}
\caption{The word cloud represents the words with the most negative impact on sentiment analysis in the context of language proficiency.}
\label{fig: n_lp}}
\qquad
\begin{minipage}{6cm}
\includegraphics[width=6cm]{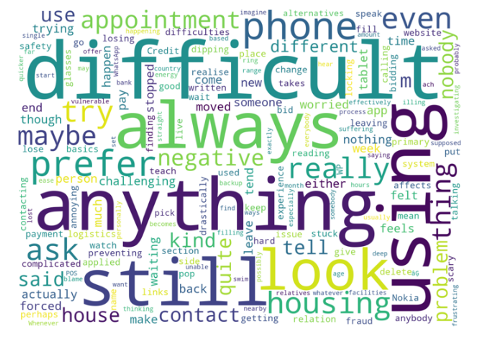}
\caption{The word cloud represents the words with the most negative impact on sentiment analysis in the context of digital literacy.}
\label{fig: n_dl}
\end{minipage}
\end{figure}

Figs. \ref{fig: n_lp} and \ref{fig: n_dl} provide a comprehensive summary of the terms that have the greatest negative impact on sentiment analysis in the digital literacy and language domains. In both figures, 'difficult' stands out as a word that explicitly reflects negative emotions in both aspects. This word conveys a sense of difficulty in relation to digital literacy and language confronted by ME individuals.



Focusing specifically on the context of language, Fig. \ref{fig: n_lp} demonstrates that "speaking" is highlighted. This suggests that difficulties with verbal communication or self-expression may negatively affect sentiment analysis in the language domain. Fig. \ref{fig: n_dl} depicts numerous digital service-related terms, including "phone," "website," "app," and "appointment." Meanwhile, the words "forced," "fraud," and "safety", which represent instances in which users encountered obstacles or experienced annoyance with digital services, contributing to the overall negative sentiment.

While Fig. \ref{fig: n_lp}  and Fig. \ref{fig: n_dl}  highlight words with a significant impact on sentiment analysis, it is important to note that not all words depicted have direct significance. In the context of digital literacy and language proficiency, words such as "try," "someone," and "come" may not contribute substantially to sentiment analysis results. Interview transcripts, a type of oral representations, present a unique set of challenges for sentiment analysis when compared to more formal sources such as news, reviews, and articles. Oral communication complicates the analysis procedure, making it more difficult to capture key words accurately.

\subsection{Verification of results using survey data with latent class analysis (LCA)}\label{lca}
To verify our findings, we applied LCA on two datasets: Evidence for Equality National Survey (EVENS)~\cite{byrne2023EVENS} which is a publicly available dataset and our own \PRIMEprojects survey. 

\begin{figure}[h]
    \centering
    \includegraphics[width=0.95\linewidth]{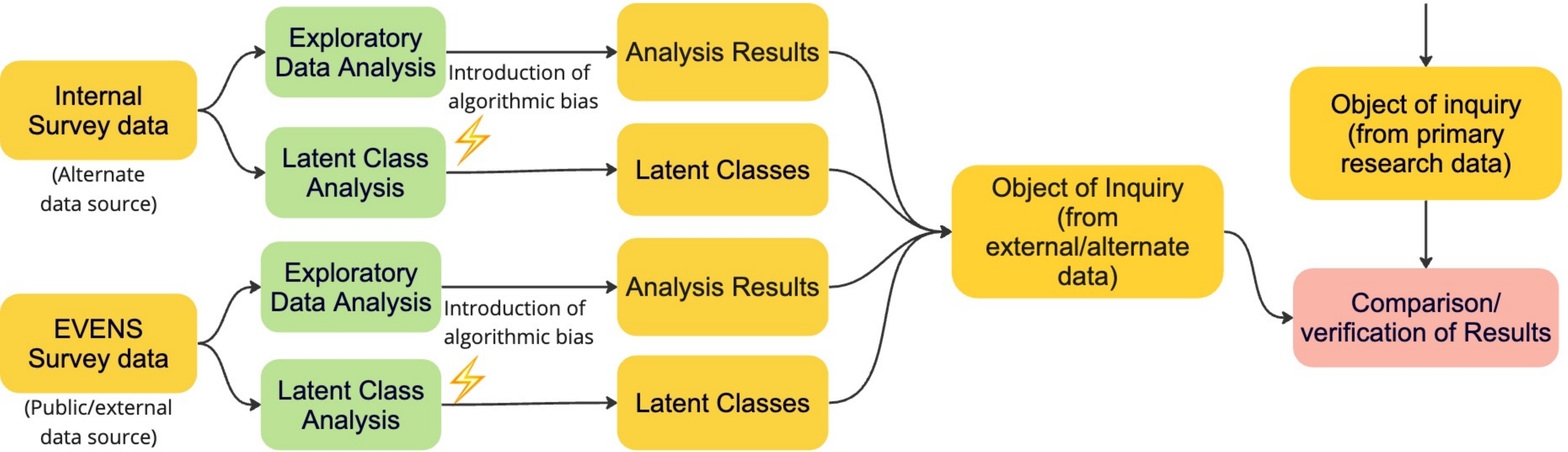}
    \caption{Latent Class Analysis done on \PRIMEprojects survey and EVENS}
    \label{fig:ML_LCA}
\end{figure}

\subsection*{\PRIMEprojects Survey}
Within the \PRIME project, a multilingual online survey (in 10 languages) was conducted with the aim of investigating the experiences of individuals from minoritised ethnic groups with the digitalised services of social housing, health, and energy.


\begin{table}[H]
\centering
\begin{tabular}{l rrr | rrr}
\toprule
\multirow{2}{*}{Ethnic Group} & \multicolumn{3}{c|}{\PRIMEproject Survey} & \multicolumn{3}{c}{EVENS Survey*} \\
\cmidrule(lr){2-7} 
& England & Scotland & Total 
& England & Scotland & Total \\
\midrule
African                         & 176 & 37  & 213       & 939  & 59  & 998 \\
Bangladeshi                     & 97  & 41  & 138    & 356  & 10  & 366 \\
Indian                          & 93  & 40  & 133       & 1194 & 47  & 1241\\
Chinese                         & 63  & 39  & 102       & 549  & 75  & 624 \\
Pakistani                       & 62  & 40  & 102      & 771  & 57  & 828 \\
Caribbean                       & 47  & 32  & 79       & 539  & 5   & 544 \\
Mixed/multiple & 56  & 55  & 111 &       &     &     \\
\midrule[1.5pt]
\textbf{Total}                  & \textbf{594} & \textbf{284} & \textbf{878} 
& \textbf{4348} & \textbf{253} & \textbf{4601} \\
\bottomrule
\end{tabular}
\caption{The number of respondents in the \PRIMEproject and EVENS surveys from England and Scotland.\\
*A subset of EVENS data was used to match ME groups investigated in \PRIMEproject survey, residing in England and Scotland only.}
\label{table:both_survey_numbers}
\end{table}

In terms of our target respondents to this survey, in England, we calculated the target number of respondents from each ethnic group proportionally, referring to their population percentages published in the 2021 Census. In contrast, as the Scotland 2021 census results were not available at the time of analysis, we aimed to set 40 as the max number of respondents for each ME groups. 
It's also important to note that the Scottish and English census categorise ethnicities differently. For example, in the census in England, ``Pakistani'' falls  under ``Asian or Asian British''. Whereas in Scotland, this equates to ``Pakistani, Pakistani Scottish or Pakistani British''. For convenience, this paper uses the ethnicity categories as in the England's census. The final tally of respondents was 878, with 594 from England and 284 from Scotland. 
Details of the respondents to the \PRIMEprojects survey and their ethnicities are shown in Table \ref{table:both_survey_numbers}, left. 



As the \PRIMEproject aims to investigate the experiences of individuals from minoritised ethnic 
groups in accessing digitalised services, the most intuitive and straightforward approach used to analyse survey data was exploratory data analysis (EDA)~\cite{tukey1977exploratory}. The four selected questions, used in this analysis, focused on 
the type of accommodation the respondents are living in, previous experience of looking for social housing or interest in looking for social housing, previous or current experience of using digital service for housing-related activities and lastly, concerns regarding the use of such digital services. 

\begin{figure}[htb]
    \centering
    \includegraphics[width=0.95\linewidth]{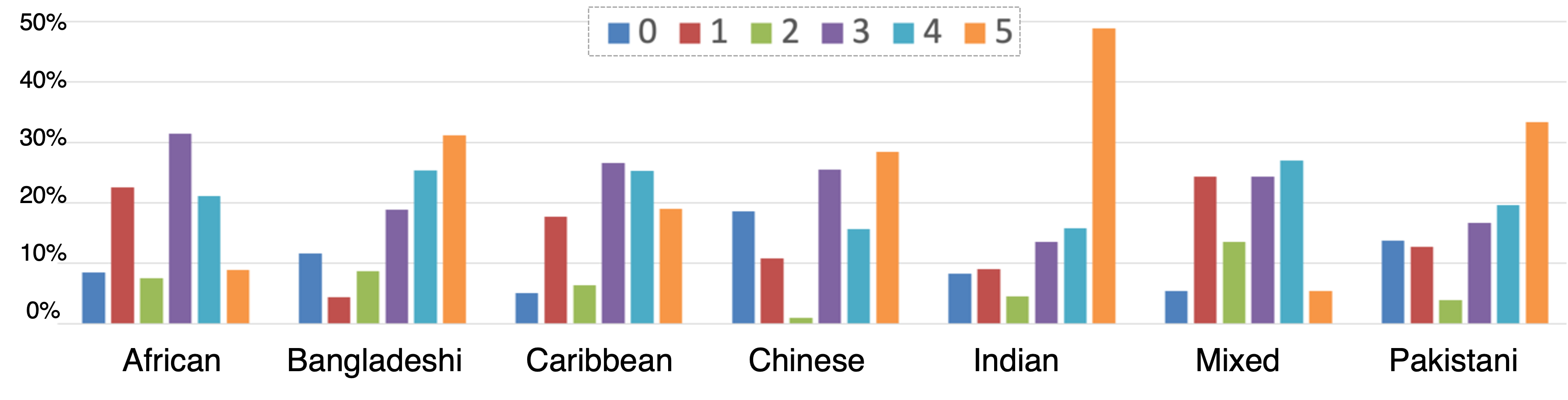}
    \caption{LCA results based on questions related to social housing, digital services and concerns, from \PRIME survey: the generated latent classes (0-5) and percentage of each ME group belonging to these latent classes. }
    \label{fig: lca_eng_scot}
\end{figure}

As the \PRIMEproject focuses on cross-sector analysis within multiple ME groups, conducting standard analysis becomes challenging when dealing with multidimensional data (i.e., when we want to analyse more than two questions simultaneously). Therefore, we employed LCA for the four aforementioned questions. The advantage of LCA is that it considers all responses as a joint distribution, unlike Exploratory Data Analysis (EDA), which analyses each question independently. The results obtained from LCA are presented in Figure \ref{fig: lca_eng_scot}.

In Figure \ref{fig: lca_eng_scot}, all participants are grouped into six different categories, determined by hyperparameter optimisation. We then visualise the percentages of participants from different ME groups within these latent categories. Overall, the majority of participants from Bangladeshi, Chinese, Indian and Pakistani backgrounds are classified into Group 5. This group represents the user profile in which individuals claim ownership of their home and show no interest in seeking social housing. Meanwhile, there are two large portions of African and Caribbean ethnicity groups classified into Group 3 which comprises of people who are not interested in looking for social housing in general due to various reasons. They may be renting privately, living with parents, living in a shared accommodation, or already have social housing available. The parameters of LCA measurement model are given in Table \ref{table: lca_prime_param} in \Cref{appendix:lca}.

\subsection*{Evidence from Equality National Survey (EVENS)}
During 2020 to 2022, the Centre on the Dynamics of Ethnicity (CoDE) 
carried out Evidence for Equality National Survey (EVENS) which documents the lives of ethnic and religious minorities in Britain during the coronavirus pandemic and is to date the largest and most comprehensive survey to do so.
In order to compare results better, in the EVENS survey dataset, we only focused on the ME groups already investigated in \PRIMEproject survey and lived in England and Scotland. This left us with 4,601 data points in total from the EVENS survey dataset. See Table \ref{table:both_survey_numbers}. 


Together with the social scientists we selected three relevant questions to conduct experiments, we also performed EDA on those three questions for England participants.
The results show that, in terms of basic living conditions, there are large differences between different ME groups. For example, Indians and Pakistanis both live more in their own homes, but others more often rent properties. Second, with respect to the two questions of whether they are concerned about being harassed and their future financial situation, all ME groups maintain a high degree of consistency in their responses, but the Bangladeshi group is the most worried about being harmed.  

\begin{figure}[ht]
    \centering
    \includegraphics[width=0.95\linewidth]{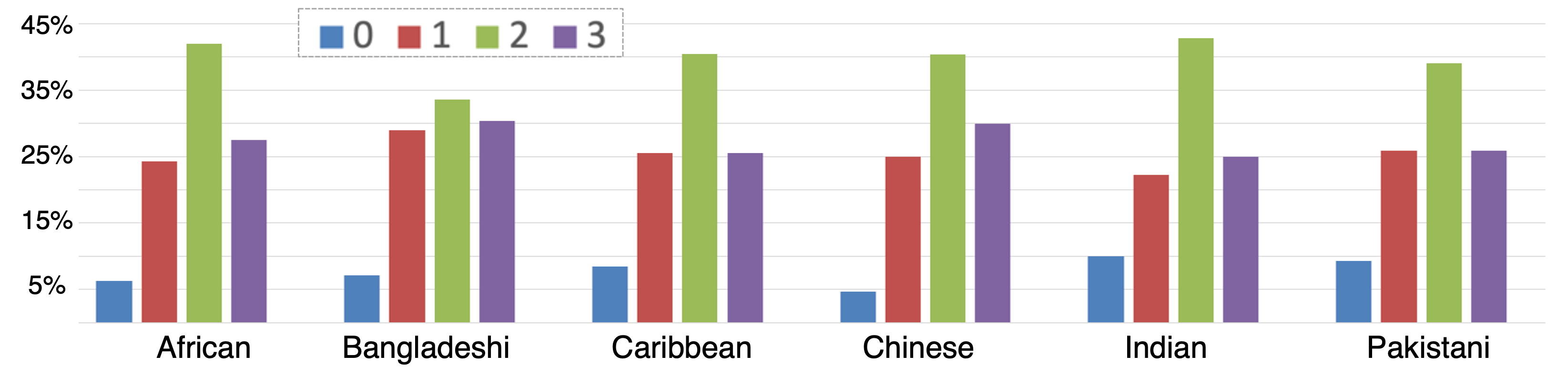}
    \caption{The result of LCA for EVENS Data: The generated latent classes (0-3) and percentage of each ME group belonging to these latent classes} 
    \label{fig: lcaevensen}
\end{figure}

Similarly, we also applied LCA on the EVENS dataset for analysis. The results are shown in Fig.~\ref{fig: lcaevensen}. From the figures, we can clearly observe that in the EVENS dataset, the Caribbean, Indian and Pakistani groups have similar clustering results. In addition, the Chinese and African groups also have similar clustering results. The experience of the Bangladeshi group comes up as unique, and not similar to that of any other ME groups. The details of the parameters of the LCA models are shown in Table \ref{table: lca_evens} in \Cref{appendix:lca}.

Comparing LCA performed on \PRIME project's survey with that on EVENS, we see a difference in results. One explanation of this could be that \PRIMEproject survey dataset was much smaller than EVENS. Also, the cross-sectoral questions we selected in EVENS are different from those in the \PRIMEproject experiment. However, it is important to note that both results show that there are clear differences in clustering results between different ME groups. This confirms that the experiences of different ME groups vary and overall, ethnic minorities cannot be treated as one homogeneous group.

\subsection{Discussion}


\textcolor{black}{
Comparing the results from the multidisciplinary analysis employing IC coupled with topic modelling and sentiment analysis with XAI, helped us paint a better picture of the state of ME experiences in the use of digitalised social housing services. The findings were then compared with the analysis of the \PRIMEproject survey and the Evidence from Equality National Survey (EVENS), further strengthening our understanding. 
}

\textcolor{black}{In particular, the analysis of interview transcripts via Topic Modelling techniques such as BiTerm, BERTopic, and CorEx helped us understand which topics the interviewees focused on, and whether certain ME groups shared the same focus. Out of the three topic modelling techniques, results from CorEx were taken forward. Although, the interviews initially focused on energy, health and social housing, emphasising on the experiences of using the digitalised form of these services, we focused our experiment with using CorEx on social housing and online experiences. The resulting distribution of topics over different ME groups 
(see Fig.~\ref{fig:topicmodelling_percentages}) show that the Black African ME groups have the highest level of relevance to the topic of social housing (29.96\%), with other groups not trailing far behind except the Chinese ME groups with the lowest level of relevance to the topic of social housing (3.72\%). This indicates different needs, attempts to access, and/or experiences of using social housing as a digitalised service. The analysis also identifies that Black Caribbean and Indian ME groups have higher levels of access and use of digitalised services in general without specific indication of which type of service (47.83\% and 44.32\% respectively). This is while the Chinese ME group shows the lowest level of relevance to digitalised services (25.40\%).
}

\textcolor{black}{As this topic modelling stage does not indicate the actual sentiment of these experiences nor the vulnerabilities of these groups, an analysis combining Sentiment Analysis with XAI was conducted. 
The sentiment analysis performed on the 21 sub-sample of interviews helped highlight the negative and/or positive sentiment. This was driven by the hypothesis that strong negative sentiment represent ME vulnerabilities. To further visualise such sentiment and vulnerabilities, XAI in the form of SHAP analysis was done. This analysis extracted words from each negative text, and those were then aggregated to generate word clouds that visually represent the negative terms in an intuitive manner.}

\textcolor{black}{Seen together, the results from CorEx topic modelling and sentiment analysis, and SHAP world clouds helped the researchers piece together the ME vulnerabilities and the barriers they face when accessing digitalised social housing services.}

\textcolor{black}{In parallel to triangulating the findings, additional data sources i.e., a publicly available survey dataset (EVENS) and an internally done survey (\PRIMEproject survey) were analysed using EDA and LCA.
However, when the results from the aforementioned ML analysis were compared with those of the both surveys, a different picture emerged. }

\textcolor{black}{For the \PRIMEproject survey, LCA grouped all participants into six different categories, determined through hyperparameter optimisation. Overall, the majority of participants from Bangladeshi, Chinese, Indian, and Pakistani backgrounds are classified into Group 5, representing the user profile where individuals claim ownership of their home and show no interest in seeking social housing. Meanwhile, there are two large portions of Black African and Black Caribbean groups classified under Group 3. This group consists of people who are either homeowners or rent privately.  For EVENS, LCA grouped all participants into 4 categories determined by hyperparameter optimisation. Here, similar to \PRIMEproject survey, we can observe that different ME groups have different experiences and, therefore, cannot be treated as one uniform group with similar conditions, vulnerabilities and experiences. }

\textcolor{black}{The overall comparison of results helped the researchers understand the varying vulnerabilities, fears, sentiments, and diversity of experiences of various ME group  in accessing digitalised social housing services. Furthermore, on the basis of the results, the researchers deemed it necessary to conduct a second round of more focused interviews to further investigate these vulnerabilities. In addition, the results supported the researchers to build on the ME perspective in the followup co-design workshops with the other stakeholders, such as service providers, policy makers and developers, as a continuity of \PRIMEproject research activities. Such workshops will focus on creating design concepts for ME harm-reducing tools for accessing digitalised social housing services.
}



\section{Conclusions}

\textcolor{black}{The paper aimed to bring together the disciplines of social science (SS) and computer science (CS) in the design and implementation of a novel unifying bias-aware 
multidisciplinary framework to systematically investigate complicated socio-technical phenomena. It proposes an innovative framework for identifying and quantifying objects of inquiry by combining analysis techniques from social sciences with those from machine learning, resulting in an novel methodological approach to translate raw qualitative data to ML friendly inputs.}

{
A case study was also presented with the aim to apply this framework in the context of minoritised ethnic (ME) experiences with digitalised social housing services. We drew our findings from 100 semi-structured interviews with ME individuals in \fourcities 
across the UK for extracting vulnerabilities. \textcolor{black}{A sub-sample of 21 interview transcripts was inductively coded to identify the topics of discrimination, digital poverty, lack of digital literacy and lack of proficiency in English in the access, use and outcomes of digitalised social housing services. The resulting data was analysed via ML techniques of topic modelling and sentiment analysis with explainable AI.} During this, the results from ML techniques were compared, fine-tuned and re-evaluated in a multidisciplinary manner. In every step, care was taken that all participating researchers were bias-aware, i.e., they took care in highlighting and understanding different sources of bias such as personal bias, algorithmic bias, classification bias and coder-dependent bias, while making informed decisions based on the results of the analysis.}

\textcolor{black}{
To further verify the results, analysis on alternative sources of data was done with the same object of inquiry in mind i.e.,  minoritised ethnic (ME) experiences with digitalised social housing services. These sources included datasets based on the \PRIMEproject survey with a part-focus on social housing, as well as Evidence for Equality National Survey (EVENS).}

\textcolor{black}{
In this case study, the way the framework was utilised in the context of interview-transcripts data focusing on minoritised ethnic (ME) experiences with digitalised social housing services was only one of many ways of implementing the framework. The framework offers the flexibility to include various combinations of different machine learning algorithms and their variants, in different ways. For data collection, interview transcripts were used as the primary source of data. However, other types of data collection methods such as large-scale surveys, questionnaires, sensor data, audio-visual data and others can also be used. Furthermore, multiple data sources can be employed to further strengthen the analysis. This may be dependent on the application field as well as the object of inquiry. }

\section{Acknowledgements}
This work is supported by \grant
\textcolor{black}{The authors would like to thank the wider \PRIMEproject team who contributed to the data collection.}

\bibliographystyle{ACM-Reference-Format}
\bibliography{main.bib}

\clearpage
\newpage
\appendix
\section*{Appendix}

\section{Parameters for Machine Learning Models}
\label{appendix:lca}
\begin{table}[htb]
\centering
\resizebox{\textwidth}{!}{
\begin{tabular}{c!{\vrule width \heavyrulewidth}rrrrrr} 
\toprule
variable                                                                                                               & \multicolumn{1}{c}{0} & \multicolumn{1}{c}{1} & \multicolumn{1}{c}{2} & \multicolumn{1}{c}{3} & \multicolumn{1}{c}{4} & \multicolumn{1}{c}{5}  \\ 
\toprule
Homeless\_0                                                                                                            & 0.011469              & 0                 & 0                 & 0                 & 0                 & 0                  \\
I am afraid that the information I disclose may be misused\_1                                                          & 0.092                 & 0.379002              & 0                 & 0                 & 0.119988              & 0                  \\
I am concerned that the digital system may work/discriminate against me\_1                                             & 0                 & 0.321898              & 0                 & 0                 & 0.025651              & 0                  \\
I am currently looking for social housing\_2                                                                           & 0                 & 0.352486              & 0.339415              & 0                 & 0.178837              & 0                  \\
I am not interested in looking for social housing\_2                                                                   & 0                 & 0.024184              & 0.017005              & 1.00000                     & 0.04926               & 1.00000                      \\
I cannot use the system due to lack of proficiency in English\_1                                                       & 0.045816              & 0.152343              & 0                 & 0                 & 0.060739              & 0                  \\
I do not currently use an app, website, or other digital service to contact or interact with my social landlord\_3     & 0.320614              & 0.01914               & 0                 & 0                 & 0.288935              & 0                  \\
I do not have any concerns about using an app, website, or digital services for these housing-related activities\_1    & 0.35895               & 0                 & 1.00000                     & 0                 & 0.511494              & 0                  \\
I do not trust my landlord\_1                                                                                          & 0.017073              & 0.141626              & 0                 & 0                 & 0              & 0                  \\
I do not understand how to use the online system\_1                                                                    & 0.114469              & 0.216772              & 0                 & 0                 & 0.182788              & 0                  \\
I don't know\_3                                                                                                        & 0.367019              & 0                 & 0                 & 0                 & 0.060644              & 0                  \\
I don’t know\_1                                                                                                        & 0.341989              & 0.037944              & 0                 & 0                 & 0.093473              & 0                  \\
I don’t know\_2                                                                                                        & 1.00000                     & 0.087884              & 0.045881              & 0                 & 0                 & 0                  \\
I expect to be looking for social housing in the future\_2                                                             & 0                 & 0.198099              & 0.071641              & 0                 & 0.542965              & 0                  \\
I have not used an app, website, or other digital service to contact or interact with a social landlord in the past\_3 & 0.204556              & 0.035907              & 0                 & 0                 & 0.172108              & 0                  \\
I have previous experience of looking for social housing\_2                                                            & 0                 & 0.513991              & 0.661859              & 0                 & 0.320798              & 0                  \\
I must disclose personal information to others who are helping me\_1                                                   & 0.079073              & 0.240286              & 0                 & 0                 & 0.066168              & 0                  \\
It takes too long\_1                                                                                                   & 0.074266              & 0.46559               & 0                 & 0                 & 0.068259              & 0                  \\
Living with parents\_0                                                                                                 & 0.112762              & 0.102458              & 0.051092              & 0.217908              & 0.110685              & 0                  \\
Other\_0                                                                                                               & 0.011469              & 0.007658              & 0.016971              & 0.02971               & 0.016267              & 0                  \\
Other\_1                                                                                                               & 0.011469              & 0.030384              & 0                 & 0                 & 0.032053              & 0                  \\
Other\_3                                                                                                               & 0.011469              & 0                 & 0                 & 0                 & 0                 & 0                  \\
Own home\_0                                                                                                            & 0.352422              & 0.106247              & 0.273079              & 0                 & 0.234717              & 1.00000                      \\
Prefer not to say\_0                                                                                                   & 0.08002               & 0.007259              & 0.017632              & 0.019816              & 0.005436              & 0                  \\
Private rented housing\_0                                                                                              & 0.227776              & 0.38611               & 0.218139              & 0.524924              & 0.348335              & 0                  \\
Shared housing\_0                                                                                                      & 0.04715               & 0.089373              & 0.068004              & 0.049479              & 0.047337              & 0                  \\
Social housing (housing rented from the council or a housing association)\_0                                           & 0.14572               & 0.212691              & 0.320975              & 0.133503              & 0.171697              & 0                  \\
Temporary accommodation\_0                                                                                             & 0.011211              & 0.088204              & 0.034109              & 0.02466               & 0.065526              & 0                  \\
Yes, to ask for advice relating to housing\_3                                                                          & 0.021924              & 0.430777              & 0.453819              & 0                 & 0                 & 0                  \\
Yes, to bid for social housing\_3                                                                                      & 0                 & 0.370188              & 0.560034              & 0                 & 0              & 0                  \\
Yes, to check rent balance and/or pay rent\_3                                                                          & 0.077933              & 0.410289              & 0.537639              & 0                 & 0                 & 0                  \\
Yes, to report the need for repairs or improvement in living conditions\_3                                             & 0.04725               & 0.348726              & 0.489387              & 0                 & 0                 & 0                  \\
Yes, to subscribe to the tenants’ portal\_3                                                                            & 0                 & 0.326636              & 0.271532              & 0                 & 0                 & 0                  \\
\toprule
\end{tabular}}
\caption{The parameters of LCA measurement model for England and Scotland \PRIMEprojects survey data}
\label{table: lca_prime_param}
\end{table}

\begin{table}[htb]
\centering
\resizebox{0.8\textwidth}{!}{
\begin{tabular}{crrrr} 
\toprule
variable                                         & \multicolumn{1}{c}{0} & \multicolumn{1}{c}{1} & \multicolumn{1}{c}{2} & \multicolumn{1}{c}{3} \\
Asylum support / home office accommodation\_0 & 0.0027                & 0.0000                & 0.0000                & 0.0000                \\
Don't know\_0                                 & 0.0274                & 0.0079                & 0.0054                & 0.0081                \\
Don't know\_1                                 & 1.0000                & 0.0000                & 0.0000                & 0.0000                \\
Extremely worried\_2                          & 0.0822                & 0.2665                & 0.0707                & 0.0000                \\
Live here rent free\_0                        & 0.0438                & 0.0397                & 0.0369                & 0.0389                \\
Live with parents\_0                          & 0.0055                & 0.0018                & 0.0011                & 0.0032                \\
No\_1                                         & 0.0000                & 0.0000                & 1.0000                & 0.0000                \\
Not at all worried\_2                         & 0.2219                & 0.3601                & 0.3158                & 0.0000                \\
Other (please specify)\_0                     & 0.0000                & 0.0009                & 0.0005                & 0.0000                \\
Own outright\_0                               & 0.2110                & 0.2145                & 0.2184                & 0.1911                \\
Own with a mortgage, other finance or loan\_0 & 0.2822                & 0.3124                & 0.3506                & 0.3530                \\
Prefer not to say\_0                          & 0.0466                & 0.0203                & 0.0112                & 0.0138                \\
Prefer not to say\_2                          & 0.1342                & 0.0335                & 0.0241                & 0.0000                \\
Rent (with or without housing benefit)\_0     & 0.3808                & 0.3998                & 0.3731                & 0.3903                \\
Shared ownership / part-mortgage part-rent\_0 & 0.0000                & 0.0018                & 0.0021                & 0.0008                \\
Somewhat worried\_2                           & 0.4438                & 0.0000                & 0.4909                & 1.0000                \\
Student accommodation\_0                      & 0.0000                & 0.0009                & 0.0005                & 0.0008                \\
Very worried\_2                               & 0.1178                & 0.3398                & 0.0985                & 0.0000                \\
Yes\_1                                        & 0.0000                & 1.0000                & 0.0000                & 1.0000               \\
\bottomrule
\end{tabular}}
\caption{The parameters of LCA measurement model for EVENS Data}
\label{table: lca_evens}
\end{table}

\label{appendix:topicmodelling}
\begin{table}[H]
\centering
\resizebox{\textwidth}{!}{
\begin{tabular}{c|c|c|c|c}
\toprule
Topic Models     & Notation & Hyperparameter details                & Range           & Step Size \\ 
\hline\hline
\multirow{4}{*}{BiTerm} 
& $\alpha$ & \makecell{The parameter of the Dirichlet prior on the per-document topic distributions} & $[0.1, 1]$      & 0.1       \\
& $\beta$  & \makecell{The parameter of the Dirichlet prior  on the per-topic word distribution}       & $[0.001, 0.1]$  & 0.001     \\
& T        & Number of topics                                                                          & $[2, 30]$       & 1         \\
& L        & Biterm generation window                                                                  & $[10, 20]$      & 1         \\ 
\hline
\multirow{2}{*}{CorEx}   
& T        & Number of topics                                                                          & $[3, 30]$       & 1         \\
& $s_a$    & Strength of the anchor words                                                              & $[1, 15]$       & 1         \\ 
\hline
\multirow{3}{*}{BERTopic} 
& T        & The minimum size of a topic                                                               & $[2, 30]$       & 1         \\
& $n_n$    & The number of neighbouring sample points in UMAP                                          & $[3, 45]$       & 1         \\
& $n_d$    & The dimensionality of the embeddings after reducing in UMAP                               & $[1, 10]$       & 1         \\
\bottomrule
\end{tabular}}
\caption{The summary of the hyperparameters selected for optimisation. The selection and settings of search spaces were based on the suggestions in the original papers~\cite{yan2013biterm, gallagher2017anchored, grootendorst2022bertopic}. The search space for CorEx was much smaller than the other two, as the design of CorEx aimed to eliminate hyperparameter constraints.}
\label{tab:hj}
\end{table}

\begin{table}[hbt]
\centering
\begin{tabular}{llllll} 
\toprule
\multicolumn{2}{c}{BiTerm ~} & \multicolumn{2}{c}{CorEx} & \multicolumn{2}{c}{BERTopic }  \\
T     &           2          & T  &       4               & T  &   2                       \\
$\alpha$ &        0.2             & $s_a$ &       2               & $n_n$ &   35                       \\
$\beta$  &          0.018           &    &                      & $n_d$ &     9                     \\
L     &              12       &    &                      &    &                          \\
\bottomrule
\end{tabular}
\caption{The summary of optimised hyperparameters values. BiTerm , CorEx and BERTopic  with the best hyperparameters were respectively optimised with coherence values $-9.5508$, $-2.0025$, $-1.8126$.}
\label{tab:hpsvalues}
\end{table}

\end{document}